 \documentclass[final,3p,times,twocolumn]{elsarticle}

\usepackage{amssymb}

\biboptions{numbers,square}

\journal{Infrared Physics and Technology}

\begin{document}

\begin{frontmatter}



\title{On the emissivity of wire-grid polarizers \\ for astronomical observations at mm-wavelengths}


\author{A. Schillaci, E. Battistelli, G. D' Alessandro, P. de Bernardis, S. Masi}

\address{Dipartimento di Fisica, Universit\'a \,  di\, Roma\, La\, Sapienza \\ P.le A. Moro 2, 00185 Roma, Italy}

\begin{abstract}
We have measured, using a custom setup, the emissivity of metallic
wire-grids, suitable for polarimeters and interferometers at mm
and far infrared wavelengths. We find that the effective
emissivity of these devices is of the order of a few $\%$, depending
on  fabrication technology and aging. We discuss their use in
astronomical instruments, with special attention to Martin Puplett
Interferometers in low-background applications, like astronomical
observations of the Cosmic Microwave Background.
\end{abstract}

\begin{keyword}
Polarizers \sep Interferometers \sep Cosmic Microwave Background


\end{keyword}

\end{frontmatter}


\section{Introduction}
\label{intro}

Wire-grid (WG) polarizers are widely used at mm-wavelengths as efficient polarization analyzers in polarimeters (see e.g. \cite{Oxle04, John07, Frais11, Monc12}), or as beamsplitters in Martin-Puplett interferometers (MPI, see \cite{Mart70}) . In fact, an array of metallic wires with diameter and spacing much smaller than the wavelength performs as an almost ideal polarizer at mm wavelengths (see e.g. \cite{Houd01}), providing an almost ideal beamsplitter for MPIs. 

In the case of astronomical observations, low-background operation  is required, to observe the faintest sources.  In general, Fourier transform spectrometers (FTS) like the MPI can measure a very wide frequency range, often covering up to two decades in frequency. This is an evident advantage with respect to dispersion and Fabry-Perot spectrometers, but comes at the cost of a higher radiative background on the detector, which is constantly illuminated by radiation from the whole frequency range covered.  

For this reason the best performing instruments are cooled at cryogenic temperatures, and are operated aboard of space carriers to avoid the background and noise produced by the Earth atmosphere. A noticeable example was the COBE-FIRAS FTS \cite{Math90}, which was cooled at T=1.5K. FIRAS measured the spectrum of the Cosmic Microwave Background with outstanding precision \cite{Fixs94}, with negligible contributions from instrumental emission. 

Intermediate performance can be obtained, quickly and cheaply, using stratospheric balloons. In this case, the residual atmosphere provides a non-negligible background, and the polarimeter/MPI can be operated at room temperature, provided that its background is kept under control. This means that the emissivity of all the optical elements of the instrument has to be as low as possible, to obtain an instrument-produced background lower than the background produced by the residual atmosphere. 

\begin{figure}
\centering
\includegraphics[height=6cm]{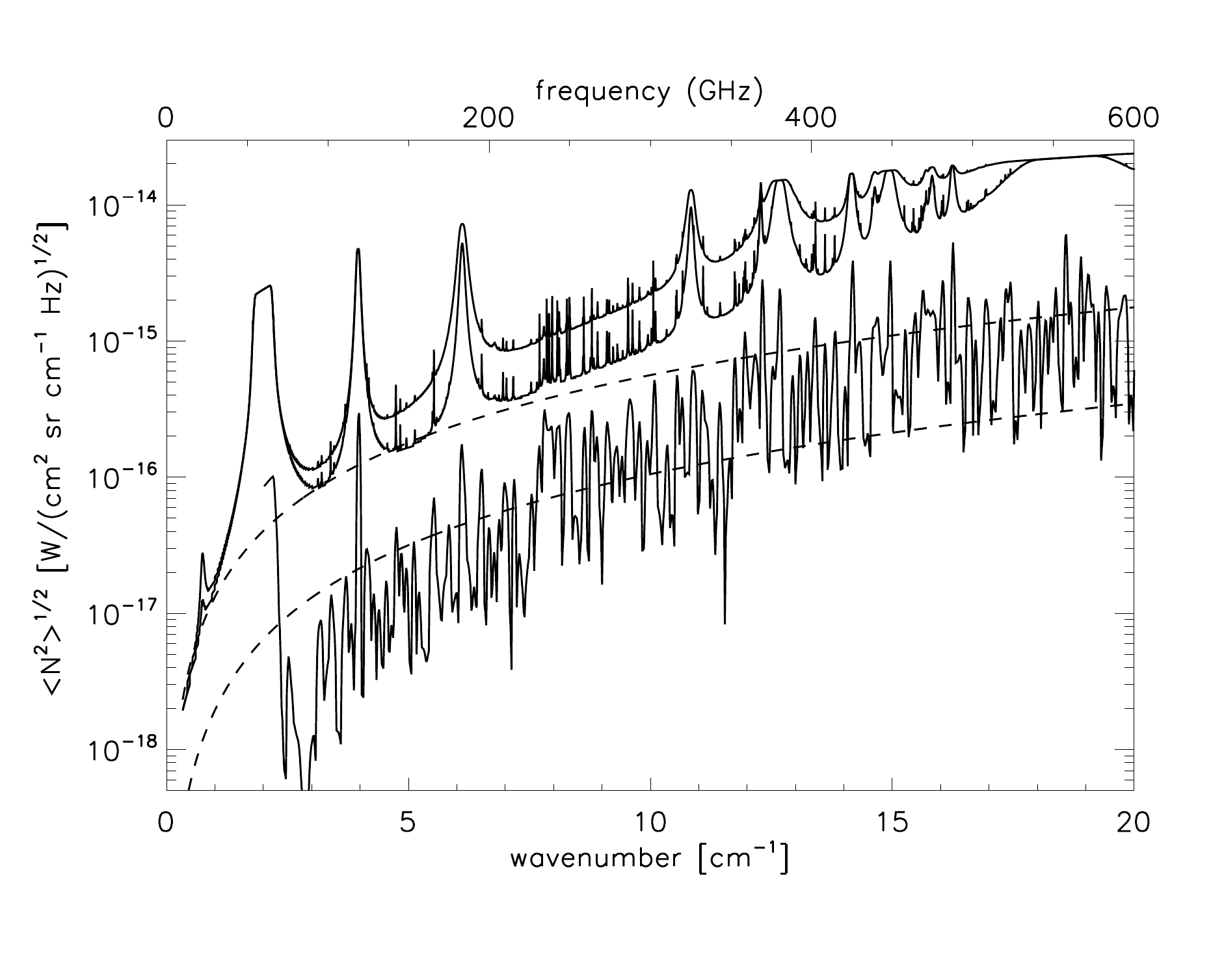}
\caption{Photon noise produced by the atmosphere, compared to the photon noise produced by a metal surface. The two top continuous lines refer to atmospheric noise in a high altitude ground based site, with 2 and 0.5 mm pwv (top to bottom); the lowest continuous line refers to photon noise produced by the residual atmosphere at stratospheric balloon altitude. In all cases we are considering a slant path to space, at 45$^o$ elevation. The two dashed lines refer to a metal surface at the same temperature as the atmosphere, with emissivity 0.02 and 0.001 (top to bottom) at $\lambda$=2 mm. 
\label{fig1} }
\end{figure}

In Figure \ref{fig1} we provide a quantitative comparison between photon noise produced by the earth atmosphere (quantum fluctuations only) and photon noise produced by low-emissivity metal surfaces (assuming a dependence on wavelength as $1/\sqrt{\lambda}$, as expected for bulk metal using the Hagen-Rubens formula \cite{Born75}). As evident from the figure, the constraint on the emissivity of the wire-grid is not very stringent for ground based observations, while it is very stringent for balloon-borne observations at mm wavelengths. 

While measurements of the emissivity of metallic mirrors in this frequency range are readily available, and for clean aluminum or brass surfaces are of the order of 0.3$\%$ at $\lambda$=2 mm (see e.g. \cite{Bock95}), the emissivity of metallic wire-grids has not been measured systematically. 

\begin{figure}
\centering
\includegraphics[height=6cm]{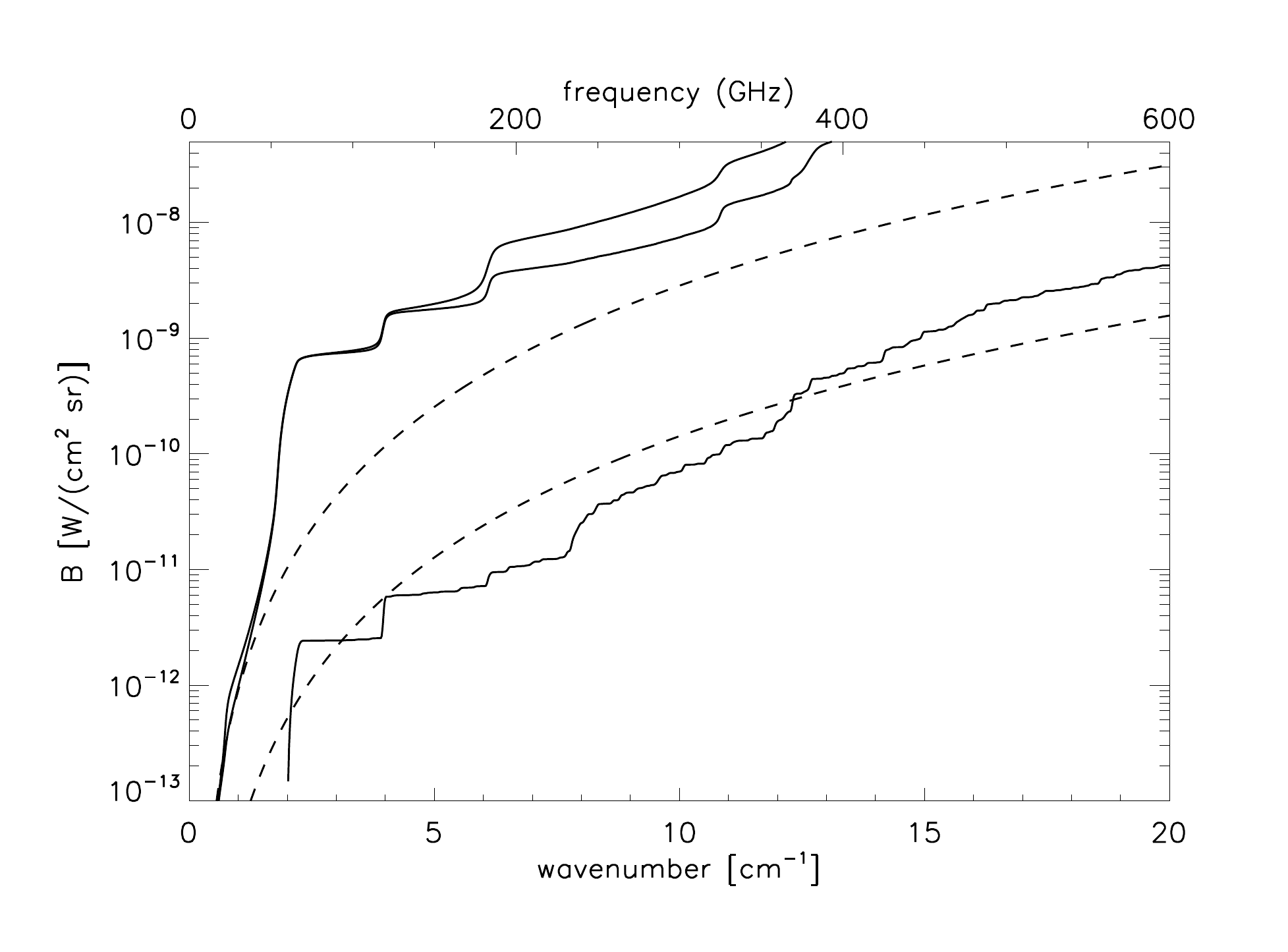}
\caption{Integral of the background power produced by the atmosphere, compared to the integral of the background power produced by a metal reflector surface. The two top continuous lines refer (top to bottom) to atmospheric emission in a ground based high altitude site, with 2 and 0.5 mm pwv; the bottom continuous line refers to emission produced by the residual atmosphere at balloon altitude (40 km). In all cases we are considering a slant path to space, at 45$^o$ elevation. The two dashed lines refer to a metal reflector surface at the same temperature as the atmosphere, with emissivity 0.02 and 0.001 (top to bottom) at $\lambda$=2 mm. 
\label{fig2} }
\end{figure}

Photon noise is not the only problem induced by emissive optical components. The average level of background radiation can saturate sensitive detectors, or seriously limit their responsivity. For this reason in Figure \ref{fig2} we provide a quantitative comparison similar to fig.1 but based on the integrated background power, over the range from 15 GHz to the frequency of interest. 

The background power values from this figure must be used to compute the background power over the frequency coverage of the instrument, and compared to the saturation power of the detector. In the case of TES bolometers for ground based measurements in the mm atmospheric windows, bolometers are designed for a saturation power ranging from $<$ 1 pW (balloon-borne experiments, low frequencies) to $>$ 10 nW (ground based experiments, high frequencies). 

At variance with metallic mirrors, where the surface can be cleaned, and high conductivity bulky metal pieces can be used, wire grids are built either with thin free-standing tungsten wires or with gold strips evaporated on a thin dielectric substrate. In both cases we do not expect that the effective emissivity is the same as for bulk metal. And we also expect that aging and oxidization can be an important effect, increasing  the emissivity of the device with time. 

From the discussion around figs. \ref{fig1} and \ref{fig2} and from the considerations above, it is evident that reliable measurements of wire-grid emissivity are in order to properly design sensitive polarimeters and MPIs for astronomical use, and decide the operation temperature of the optical components.

In this paper we describe a measurement setup we have developed to measure the effective emissivity of wire grids, at temperatures close to room temperature, at mm-wavelengths. We discuss the instrument design, report the results of measurements of different wire-grids, and discuss their application in the case of balloon-borne MPIs for mm-wave astronomy. 

\section{Measurement setup} \label{setup}

\begin{figure}
\centering
\includegraphics[width=8cm]{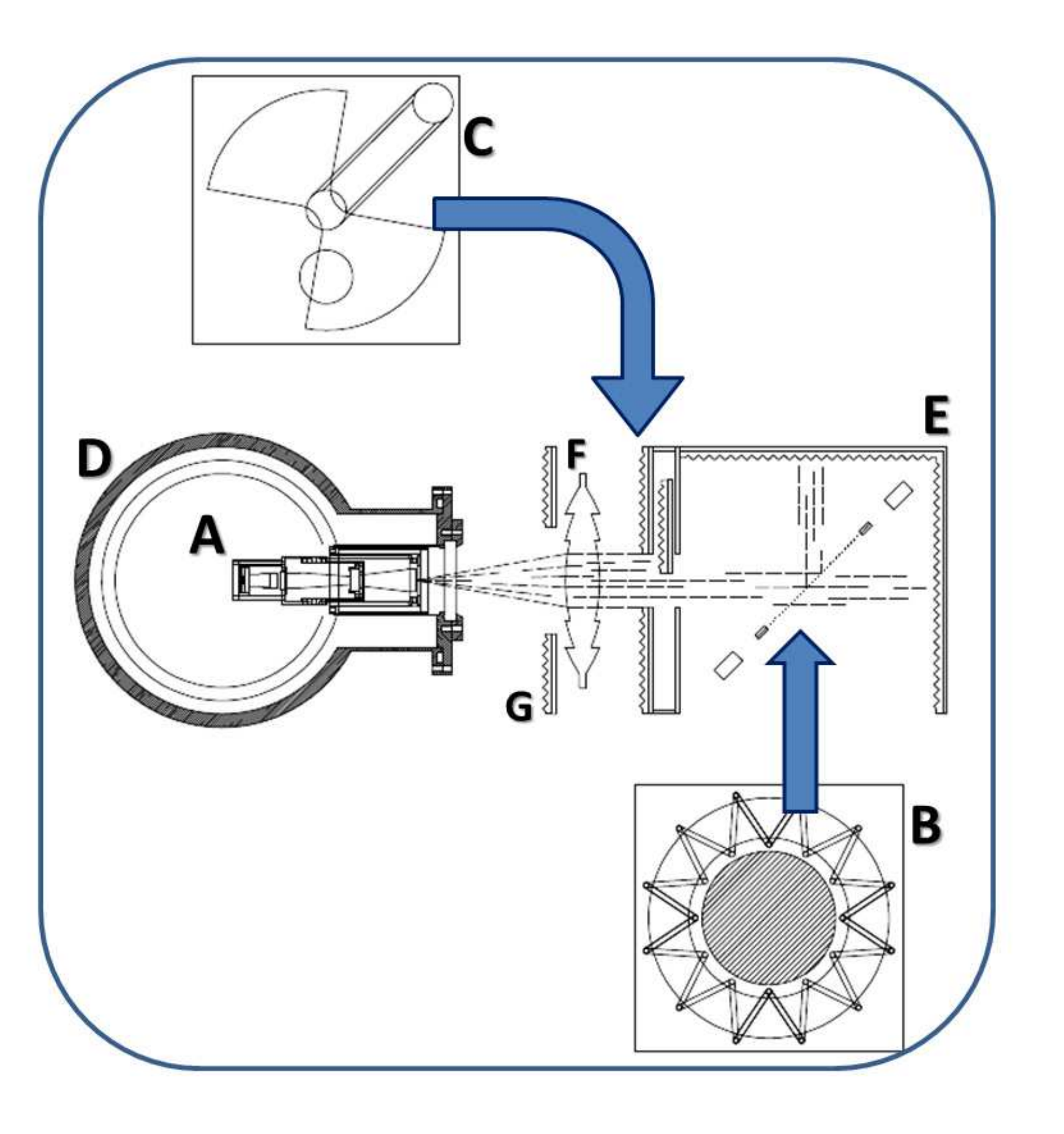}
\caption{Block diagram of the measurement setup. A : 0.3K bolometer;  B : suspended wire-grid with PID thermal control; C : eccosorb$_{tm}$ coated chopper; D : PT crycooler with $^{3}He/^{4}He$ cryostat; E : eccosorb$_{tm}$ coated cavity at room temperature; F : Fresnel lens; G:  diaphragm.
\label{sys1} }
\end{figure}

\begin{figure}
\centering
\includegraphics[width=7cm, angle=90]{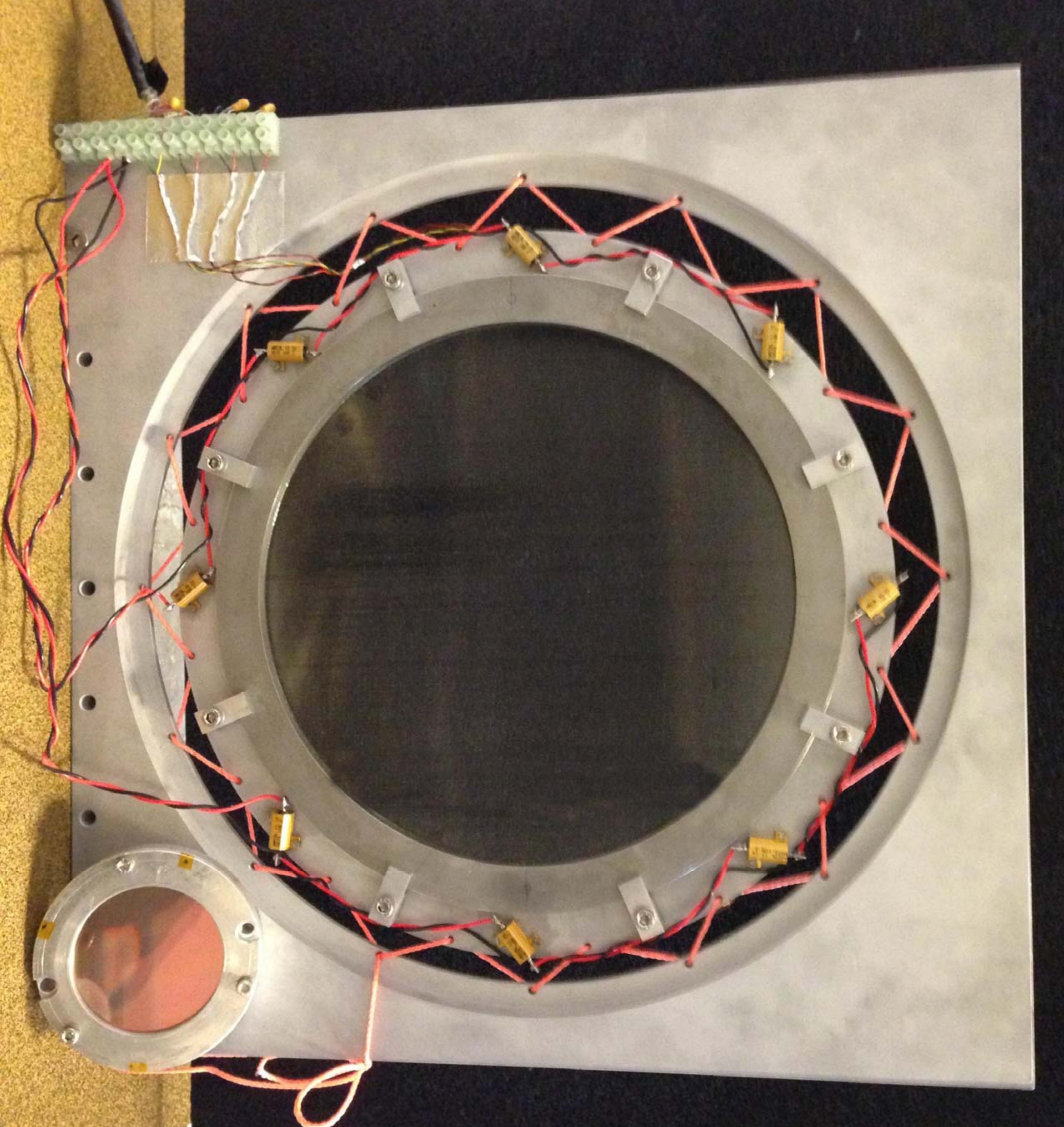}
\caption{Picture of the wire-grid with its suspension system, made with kevlar cords (1.6 mm diameter, 38 mm long)
\label{sys2} }
\end{figure}

In our approach the emissivity is measured heating the WG and detecting the change of emission.  The WG is inserted in a room-temperature blackbody cavity, with walls covered by eccosorb \emph{72AN} foils, 6 mm thick, so that both transmitted and reflected radiation are carefully controlled. The radiation emitted, transmitted and reflected by the WG is modulated by a room-temperature chopper with eccosorb-coated blades, and is detected by a 0.3K bolometer. The wire grid is mounted on a metal ring suspended in the middle of the blackbody cavity by means of kevlar cords, and can be heated by power resistors. The design of the suspension support for the WG was optimized in order to have a value for the time constant of the heating process not too high. By this way we calculate the thermal capacitance of the wire grid support and we assume that the two prevalent ways to disperse heat are the conductivity of the kevlar wires and the convection process on the wire grid surfaces. The best compromise was to design a very low profile alluminium ring with a weight of only 259 grams and 32 links of 38 millimeters lenght 1.5 millimiters in diameter kevlar wire (see fig.3). The ring is equipped with PT100 thermometers, biased by a 1 mA current and readout by a 16-bit ADC. Temperature stabilization is obtained by means of a digital PID algorithm, which uses the thermometer reading to drive a PWM current through the power resistors. We are particularly interested to the D band, so our detector bandpass ranges from 125GHz to 175GHz. The NEP of the system under room-temperature background is around $10^{-15}W/\sqrt{Hz}$.  This detector is the same used for the BRAIN experiment \cite{Masi05, Pole07, Batt12}. The detector is AC biased at 415Hz, while the chopper modulates the optical signal at 20 Hz. Two lock-in amplifiers are cascaded for demodulation. A block diagram and a picture of the system are shown in figs. \ref{sys1} and \ref{sys2}. 

We measured the emissivity of three different types of wire grid:
\begin{enumerate}
\item 23 cm diameter wire-grid, made of tungsten filaments, 10 $\mu m$ diameter, 4 years old (WG1)
\item 5 cm diameter wire-grid made of tungsten filaments, 10 $\mu m$ diameter, 9 months old (WG2)
\item 5 cm diameter wire-grid made of evaporated gold wired on polipropylene film (WG3)
\end{enumerate}

\section{Measurement Method }

When the chopper is closed the detector receives radiation from the eccosorb-coated chopper blades, at a temperature $T_ {cho}$ and with emissivity $\varepsilon_{ecc}$. When the chopper is open, the detector recives radiation emitted by the WG, with wires at physical temperature $T_ {wg}$ and with emissivity $ \varepsilon_ {wg} $, radiation transmitted by the WG and radiation reflected and/or scattered by the WG. The latter two come from  blackbodies at $T_{amb}$.  Working with thermal radiation in the Rayleigh-Jeans region, the power emitted is proportional to the physical temperature of the emitter. So we can write the measured signal, after demodulation, as:

$$ S={\cal R}[\varepsilon_{ecc}(T_{amb} (t_{wg}+r_{wg})-T_{cho})+\varepsilon_{wg}T_{wg}] $$

where ${\cal R}$ is the system responsivity in $V/K$, while $t_{wg}$ and $r_{wg}$ are the wire-grid trasmissivity and reflectivity. The product ${\cal R}\varepsilon_{ecc}$ is calibrated removing the WG and placing an eccosorb plate cooled at $T_N=77.8K$ behind the chopper: in this case, in fact, 

$$ S_{cal}={\cal R}[\varepsilon_{ecc}(T_N -T_{cho})] $$

In addition, the eccosorb emissivity can be estimated measuring the transmissivity (very low) and the reflectivity (also very low) of one of the eccosorb slabs, using a powerful source (a 150 GHz Gunn oscillator) and the same detector. We get $\varepsilon_{ecc}=0.972 \pm 0.002$. Once the responsivity $\cal R$ is known, then the ratio $S/{\cal R}$ can be estimated. This is a linear function of $T_{wg}$. The slope of the line is equal to the WG emissivity $\varepsilon_{wg}$, which can be estimated from a best fit of the $S/{\cal R}$ measured for different $T_{wg}$.

\section{Results and discussion} \label{results}

\begin{figure}
\centering
\includegraphics[width=7.5cm, angle=180]{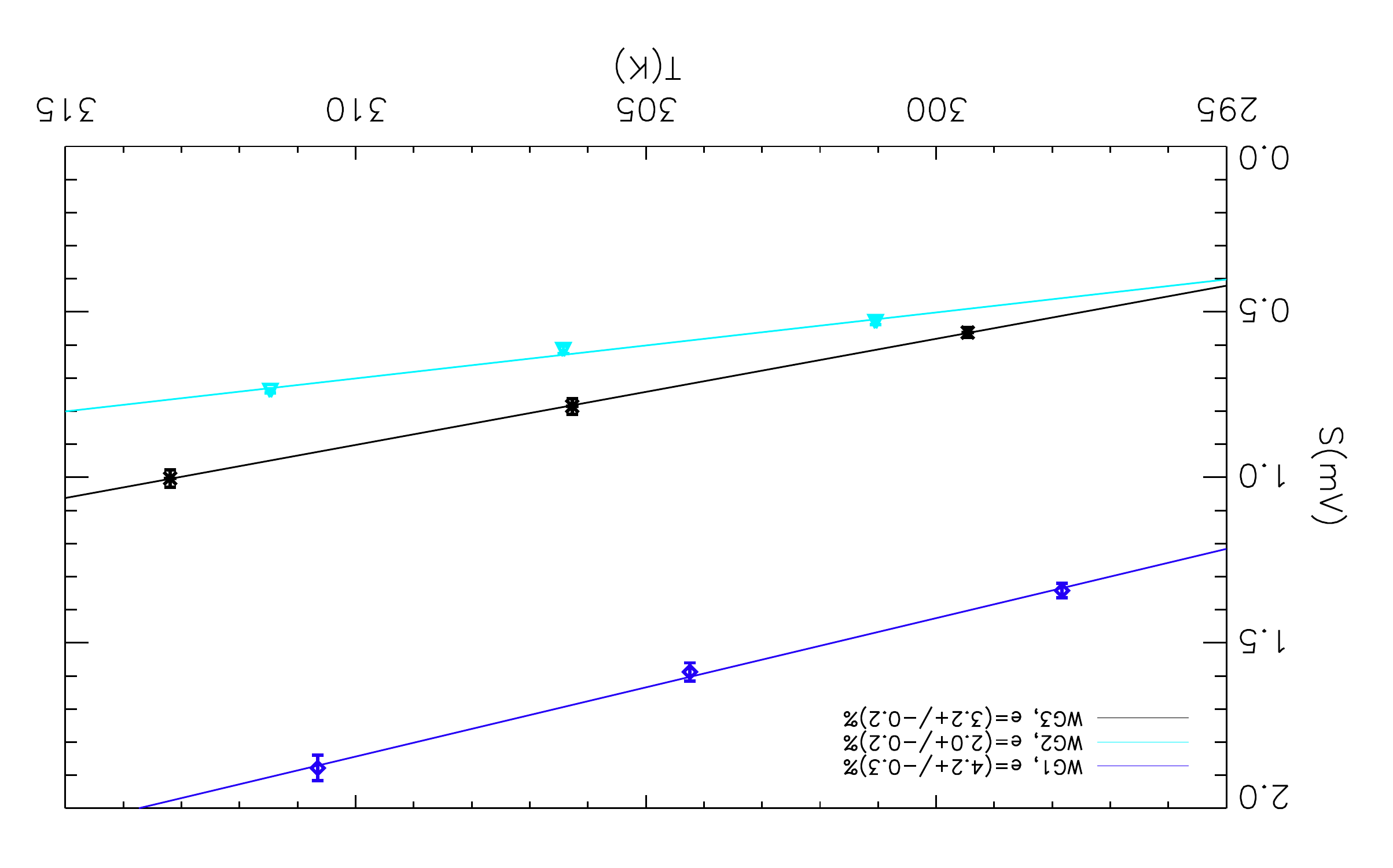}
\caption{Signals $S$ detected for different temperatures $T_{WG}$ of the WGs. The emissivities of the WGs are estimated from the slopes of the best fit lines. The symbols are diamonds for WG1, triangles for WG2, asterisks for WG3.
\label{signals} }
\end{figure}

In Figure \ref{signals} we plot the signals detected for the three wire grids. 

From the slopes of the best fit lines and the calibration described above we estimate the  emissivities reported in Table\ref{tab:emis}.

One potential systematic error source is a temperature gradient along the metal wires of the WG. These wires are heated by the metal support ring at their edges, but are cooled by ambient air at the center. Tungsten is not a very good heat conductor, and evaporated gold wires are very thin. For these reasons we investigated this issue carrying out a finite-elements thermal analysis on a model of the system. We have used Comsol Multiphysics \cite{comsol} to simulate the transmission of heat in a tungsten filament of different lengths $\ell$, in thermal contact at both edges with a thermostat at temperature $T_2$. The wire is suspended in still air at normal pressure and with temperature $T_3$, and heat exchange between the wire and air is taken into account.  Assuming the wire center temperature is $T_1=T_3$ at the beginning, its regime center temperature $T_r$ is computed. The results are given in Table \ref{tab:tempwire}.

\begin{table}
\begin{center}
\begin{tabular}{|c|c|}
  \hline
   & $\varepsilon_{wg}$ \\
  \hline
  $WG1$ & $(4.2 \pm 0.3)\%$\\
  $WG2$ & $(2.0 \pm 0.2$)\%\\
  $WG3$ & $(3.2 \pm 0.2$)\%\\
  \hline
\end{tabular}
\caption{Measured emissivity for the three wire grids described in the text.
\label{tab:emis}}
\end{center}
\end{table}

\begin{table}
\begin{center}
\begin{tabular}{|c|c|c|c|c| }
  \hline
 $\ell$(cm) &  $T_1$ & $T_2$  & $T_r$ & ${(T_2 - T_r) \over (T_2 - T_1)}$ \\
  \hline
  23 & 295 & 300,0 & 298,7 & 26\% \\
  23 & 295 & 305,0 & 302,5 & 25\% \\
  23 & 295 & 312,0 & 308,1 & 23\% \\
  5 & 295 & 300,0 & 299,8 & 4\%\\
  5 & 295 & 305,0 & 304,9 & 1\%\\
  5 & 295 & 312,0 & 311,9 & 0.6\%\\
  \hline
\end{tabular}
\caption{Estimate of the regime temperature $T_r$ at the center of a 10$\mu$m diameter tungsten wire, $\ell$(cm) long, connected at the two extremes to a thermostat at temperature $T_2$, in still air at temperature $T_1$. The last column gives the maximum relative error generated in the emissivity measurement  from assuming than the temperature of the wire is the same as the temperature of the thermostat. 
\label{tab:tempwire}}
\end{center}
\end{table}

The estimates in Table\ref{tab:tempwire} show that for the smaller diameter WGs (WG2 and WG3) the systematic error is smaller than the statistical error of our measurements. For the large WG (WG1) this systematic error can produce an underestimate of the emissivity of the order fo 10\% (taking into account that the $T_r$ is estimated in the middle of the wire, while the temperatures of the edges are correct). 

Comparing to the noise and background produced by the atmosphere plotted in fig.\ref{fig1} and fig.\ref{fig2} we see that room-temperature WGs contribute very significantly to the total background on the detector. 

In the basic design of a MPI three WGs are needed. While the background generated from the WG can be neglected in the case of ground based astronomical observations in the mm atmospheric windows, it is the dominant background in the case of balloon-borne observations with a room.temperature interferometer. 

\begin{figure}[t]
\centering
\includegraphics[scale=0.42, angle=90]{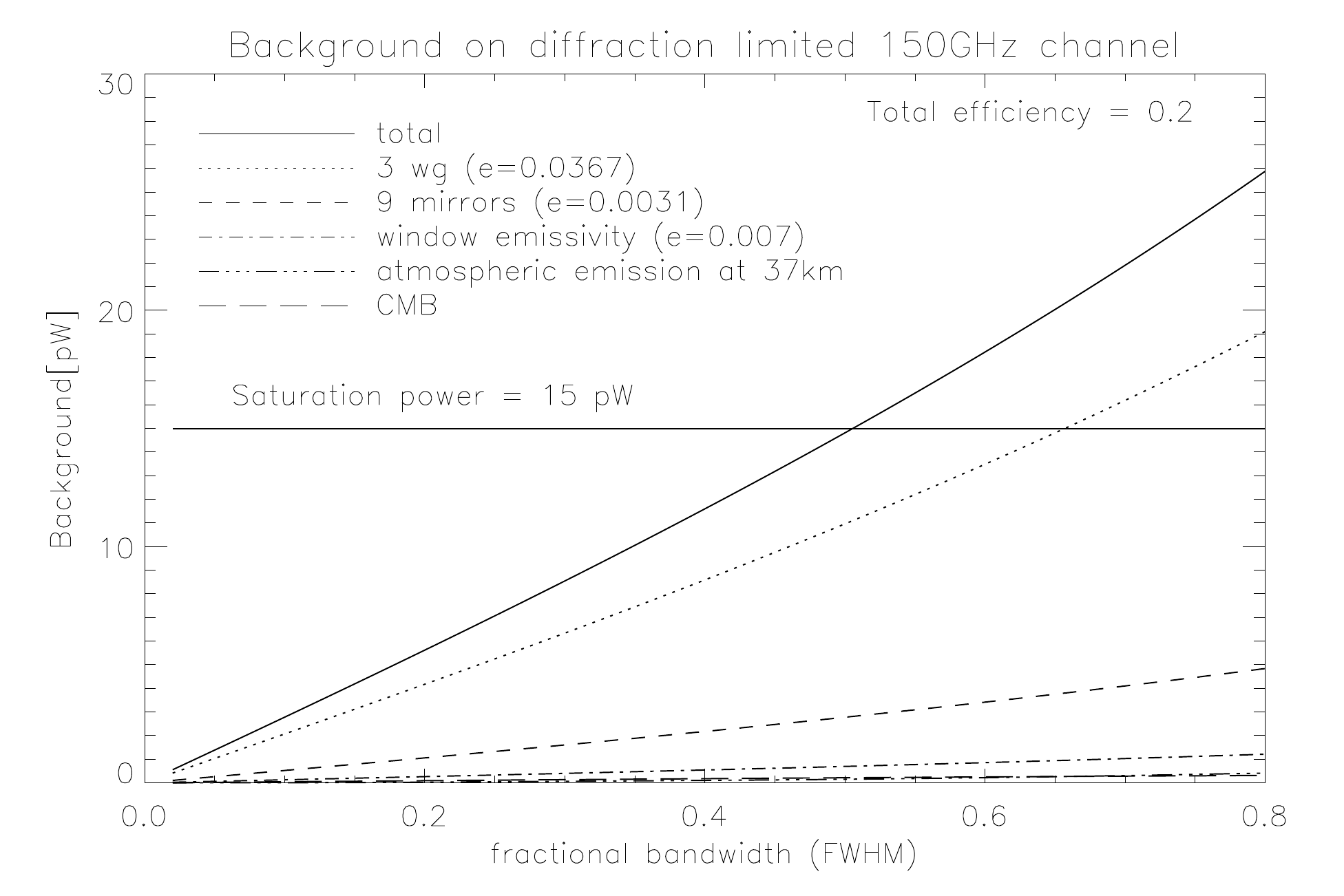}
\caption{Example of total radiative background on the detector vs bandwidth for a detector working in the 140 GHz band, in the case of a balloon-borne room-temperature spectrometer. 
\label{bkg150} }
\end{figure}

\begin{figure}[t]
\centering
\includegraphics[scale=0.42, angle=90]{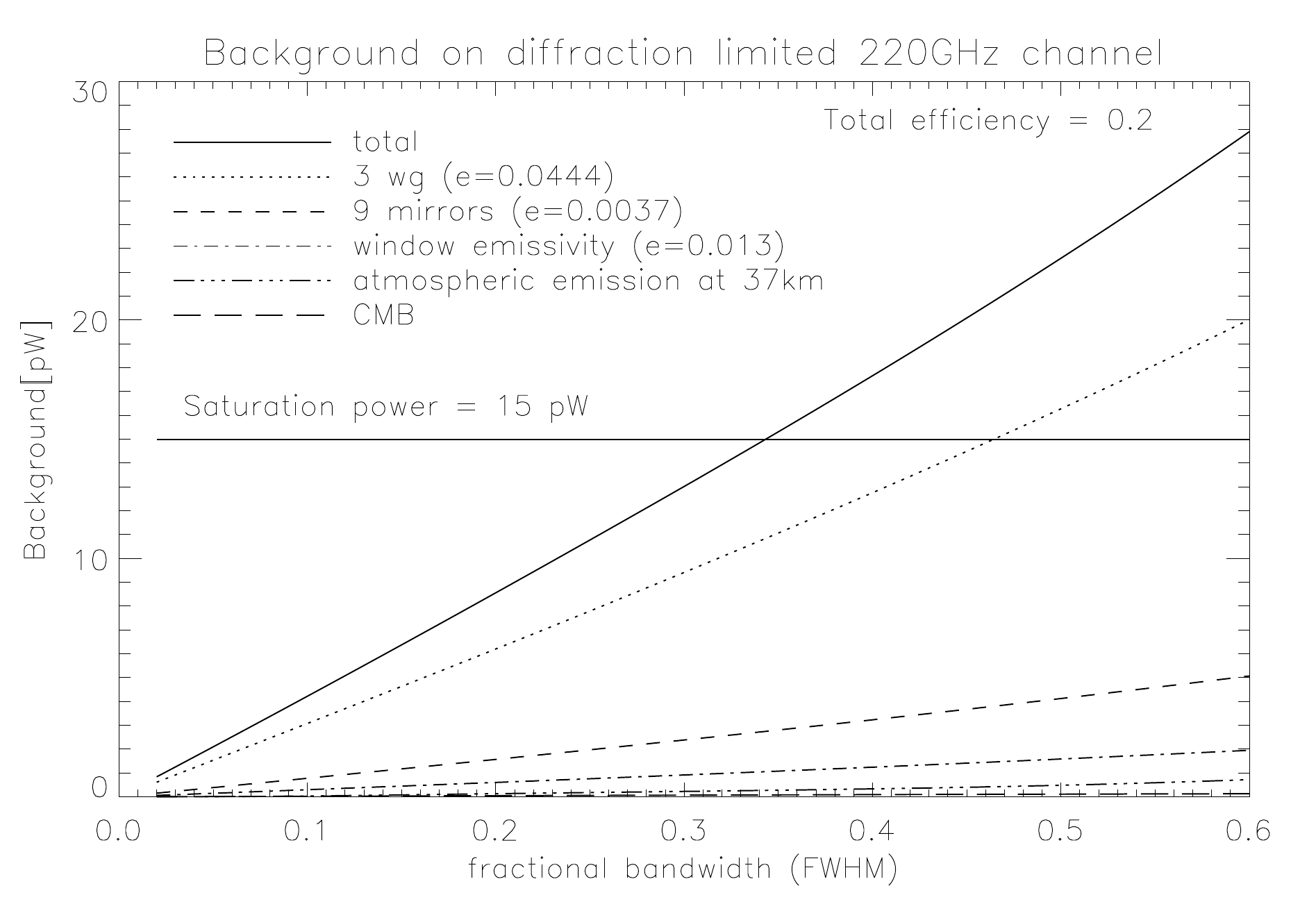}
\caption{Same as fig. \ref{bkg150} for the 220 GHz band. Here the limitation to the width of the band is very relevant.
\label{bkg220} }
\end{figure}

For example, in the case of ballon-borne observations of the CMB in the 140 and 220 GHz bands, the different contributions to the total background on the detector are summarized in figs. \ref{bkg150} and \ref{bkg220} as a function of the band width. It is evident that the presence of room-temperature WGs dominates the total background on the detectors. This is specially important in the case of TES detectors,  where this additional  background challenges the saturation power, limiting in practice the width of the band. 

\begin{figure}[t]
\centering
\includegraphics[scale=0.48]{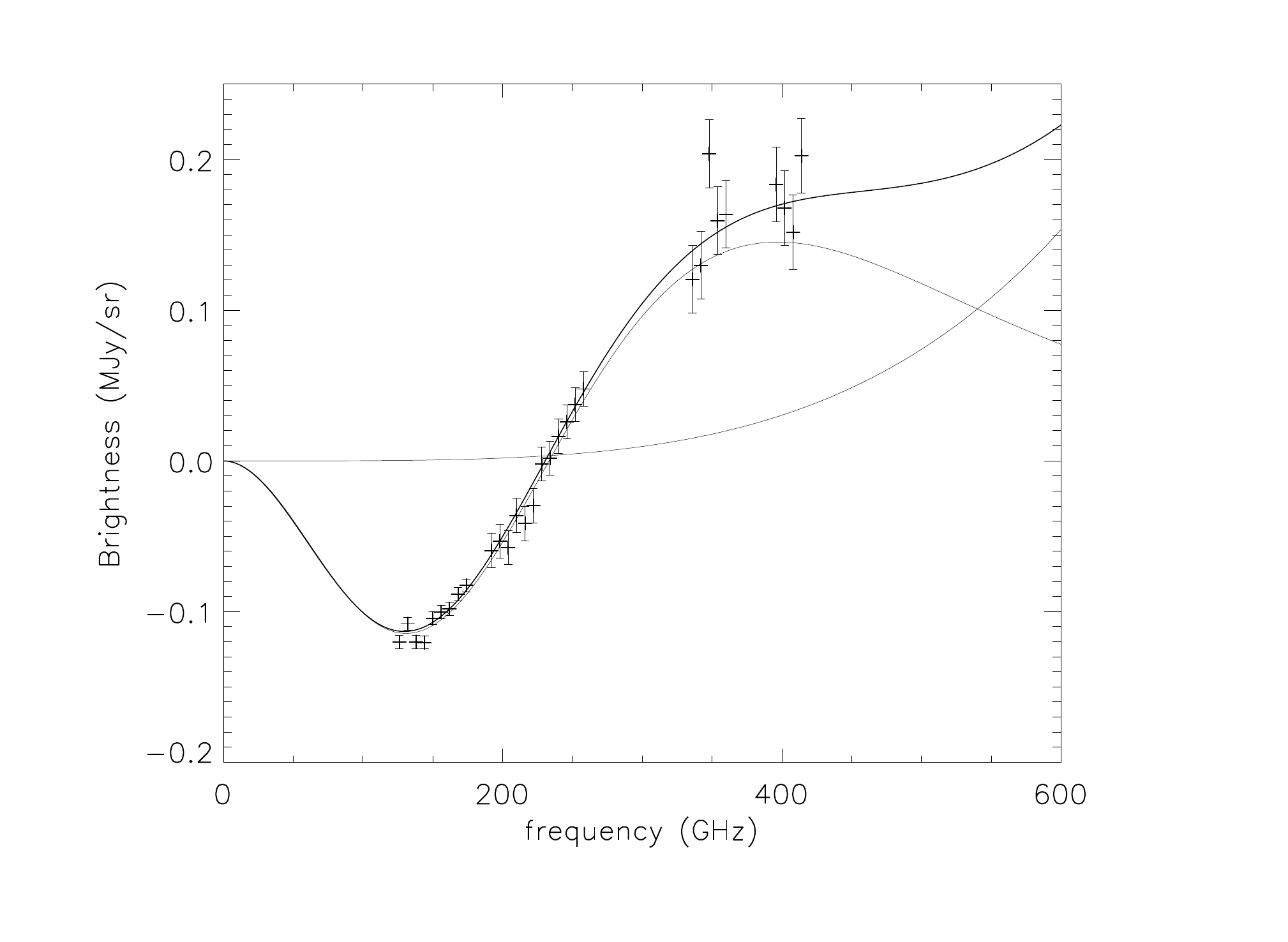}
\caption{Simulation of a 3-hours observation of the Sunyaev-Zeldovich effect in a cluster of galaxies, obtained using a room-temperature differential MPI on a stratospheric balloon, with TES detectors working in 4 bands. The detectors are photon-noise limited, and the radiative background is computed as described in figs. \ref{bkg150} and \ref{bkg220}. The thick line is the best fit to the simulated observations, and the two thin lines represent the two contributing components: SZ effect and interstellar dust emission. \label{SZ} }
\end{figure}

In fig. \ref{SZ} we plot a simulation of a 3-hours observation of the Sunyaev-Zeldovich (SZ) effect in a typical cluster of galaxies, obtained using a room-temperature differential MPI on a stratospheric balloon (like the OLIMPO experiment \cite{Masi08}), with photon-noise limited TES detectors working in bands optimized as described above. See \cite{debe12} for details. The potential of this method is evident, despite of the bandwidth limitation due to the emissivity of the WGs discussed here.

\section{Conclusions} \label{conclusion}

 We have measured the emissivity of different wire grids at 150 GHz. The emissivity, of the order of a few \%, is larger than the one of metallic mirrors normally used at these frequencies and dominates the emissivity budget of a MPI. This is relevant in the case of observations carried out with a room-temperature interferometer at balloon altitude, because it dominates the radiative background of the instrument.

\section{Acknowledgments}
This work has been supported by ASI (Agenzia Spaziale Italiana) grants OLIMPO, BOOMERanG and Millimetron, and by PNRA (Italian National Antarctic Research Program) grant BRAIN.
We warmly acknowledge Mr. Giorgio Amico for carefully machining many parts of the instrument.






\begin{thebibliography}{00}

\bibitem{Oxle04} P. Oxley, P. Ade, C. Baccigalupi, et al., “The EBEX experiment”, Proc.SPIE Int.Soc.Opt.Eng. 5543 (2004) 320-331

\bibitem{John07}  B. R. Johnson, J. Collins, M. E. Abroe, et al. “MAXIPOL: Cosmic Microwave Background Polarimetry Using a Rotating Half-Wave Plate”, Ap.J., 665, 42–54, (2007) 

\bibitem{Frais11}  A. A. Fraisse, P. A. R. Ade, M. Amiri, et al., “SPIDER: Probing the Early Universe with a Suborbital Polarimeter”, astro-ph/1106-3087, (2011)

\bibitem{Monc12}  Moncelsi, L., Ade, P., Angile, F. E., et al., “Empirical modelling of the BLASTPol achromatic half-wave plate for precision submillimetre polarimetry“, astro-ph/1208-4866 (2012)

\bibitem{Mart70} D. Martin and E. Puplett, “Polarised interferometric spectrometry for the millimetre and submillimetre spectrum”, Infrared Physics, 10, 105-109, (1970).

\bibitem{Houd01} Houde M., Akeson R. L., Carlstrom J. E., et al., “Polarizing Grids, Their Assemblies and Beams of Radiation”, Publ. Astron. Soc. Pac., 113, 622-638 (2001)

\bibitem{Math90}  Mather, J. C., Cheng, E. S., Eplee, R. E., Jr., et al., Ap.J., 354, L37-L40 (1990)
 
\bibitem{Fixs94} Fixsen, D. J., Cheng, E. S., Cottingham, D. A., et al., ApJ, 420, 445-449 (1994)

\bibitem{Born75} M. A. Born and E. Wolf, Principles of Optics, Pergamon, Oxford, pg. 611–633 (1975)

\bibitem{Bock95} J. J. Bock, M. K. Parikh, M. L. Fischer, and A. E. Lange “Emissivity measurements of reflective surfaces at near-millimeter wavelengths”, Applied Optics, 34, 4812-4816 (1995)

\bibitem{Masi05} S. Masi, P. de Bernardis, C. Giordano, et al. "Precision CMB Polarization from Dome-C: the BRAIN experiment"  EAS Publications Series, 14 , 87-92 (2005)

\bibitem{Pole07} G. Polenta, P.A.R. Ade, J. Bartlett, et al. "The BRAIN CMB polarization experiment", New Astronomy Reviews, 51, 256-259, (2007) 

\bibitem{Batt12} E. S. Battistelli, G. Amico, A. Ba\'u, et al. "Intensity and polarization of the atmospheric emission at millimetric wavelengths at Dome Concordia", Mon. Not. R. Astron. Soc. 423, 1293–1299 (2012)

\bibitem{comsol} http://www.comsol.com/products/multiphysics/

\bibitem{Masi08} Masi S., Battiatelli E., Brienza D., et al., OLIMPO, Mem. S.A.It., 79, 887 (2008)

\bibitem{debe12} P. de Bernardis, S. Colafrancesco, G. D' Alessandro, et al.,  Low-resolution spectroscopy of the Sunyaev-Zeldovich effect and estimates of cluster parameters, Astronomy and Astrophysics, 538, A86 (2012)





\end{thebibliography}


\end{document}